\documentclass[preprint,prl,unsortedaddress,superscriptaddress]{revtex4}
\usepackage{amsmath,color}
\usepackage[pdftex]{graphicx}
\allowdisplaybreaks
\begin{document}
\title{Negative Effective Density in An Acoustic Metamaterial}

\author{Sam Hyeon Lee}
\affiliation{Institute of Physics and Applied Physics, Yonsei
University, Seoul 120-749, Korea}
\author{Choon Mahn Park}
\affiliation{AEE Center, Anyang University, Anyang 430-714, Korea}
\author{Yong Mun Seo}
\affiliation{Department of Physics, Myongji University, Yongin
449-728, Korea}
\author{Zhi Guo Wang}
\affiliation{Department of Physics, Tongji University, Shanghai
200092, People's Republic of China}

\author{Chul Koo Kim\footnote{To whom correspondence should be addressed. E-mail:
ckkim@yonsei.ac.kr}}
\affiliation{Institute of Physics and Applied
Physics, Yonsei University, Seoul 120-749, Korea}

\date\today

\begin{abstract}

\textbf{Abstract.} We report theoretical and experimental results
for a new type of homogenized acoustic metamaterials with negative
effective mass density. We constructed one-dimensional metamaterial,
which is a tube with an array of very thin elastic membranes placed
inside. This structure exhibited negative effective density in the
frequency range from 0 to 735 Hz. The experimental result is in
excellent agreement with our theoretical model that predicts
negative effective density below a cut-off frequency. The frequency
characteristics of this effective density is analogous to that of
the permittivity of the electromagnetic plasma oscillation.

\end{abstract}

\maketitle

The realization of double negative electromagnetic metamaterials has
opened a new area of science in the field of electromagnetic waves
~\cite{1,2,3,4,5,6}. In the electromagnetism, the two constitutive
parameters, the electric permittivity, $\epsilon$, and magnetic
permeability, $\mu$, determine the phase velocity, $v_{ph} =
\sqrt{1/\epsilon \mu}$, in a medium. If $\epsilon$ and $\mu$ are
simultaneously negative (double negativity, DNG), the waves
propagate through the media with a phase velocity which is
antiparallel to the Poynting vector. Progress in electromagnetic
metamaterials also stimulated researches in acoustic
metamaterials~\cite{7,8,9,10,a1,a2,11}. If the density and the bulk
modulus are simultaneously negative, the phase velocity of sound is
expected to also become negative. Because no naturally occurring
material exhibits such negative parameters, acoustic DNG has to be
achieved using engineered subwavelength structures. Negative bulk
modulus has been realized using an array of Helmholtz
resonators~\cite{zhang}. As for the negative density, a single
resonator consisting of a rubber membrane with a central disk has
been demonstrated to exhibit negative dynamic mass~\cite{9}.
However, a negative mass density in a medium has not been realized
yet.

\begin{figure}
\begin{center}
\includegraphics*[width=0.5\columnwidth]{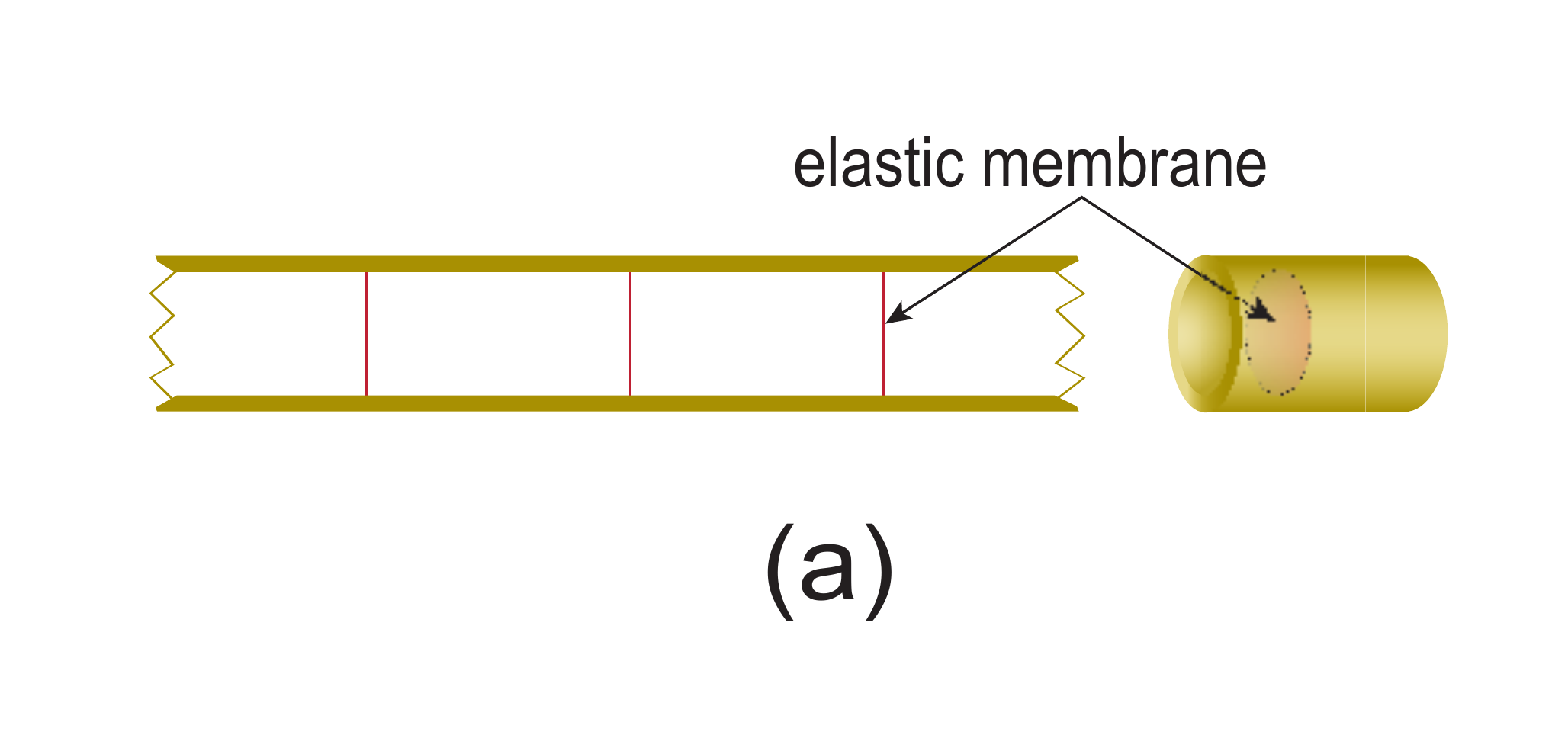}
\includegraphics*[width=0.8\columnwidth]{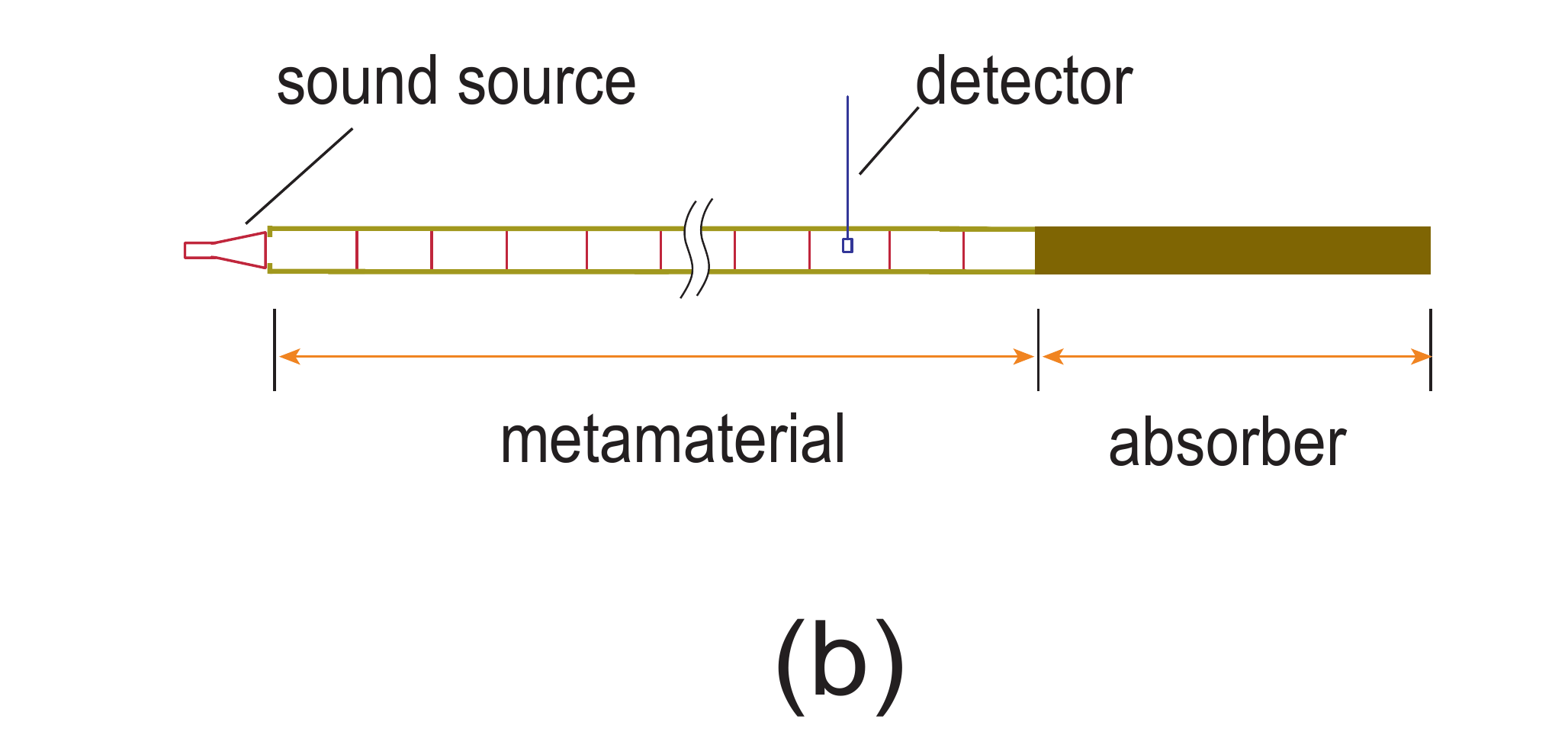}
\end{center}
\caption{(color online) (a) Structures of metamaterials;
One-dimensional structure consisting of thin tensioned elastic
membranes in a tube. Negative effective density is observed in this
system. (b) Experimental setup for the transmission and phase
velocity measurements.} \label{fig:diag}
\end{figure}

In this paper we present a new type of homogenized acoustic
metamaterial that exhibits negative effective density for a wide
range of frequencies. This metamaterial is based on a new type of
elasticity not observed in naturally occurring materials. This
elasticity is generated in a fluid by using an array of very thin
tensioned membranes. We present a calculation showing that this
elasticity creates negative effective density for all frequencies
below a cut-off frequency instead of a stop band. So far, all known
negative constituent parameters, either for electromagnetic waves or
for acoustic waves, have stop bands except for the effective
permittivity of the electromagnetic plasma oscillation. Our model
predicts the frequency characteristics of negative density exactly
of the same form as that of the effective permittivity in a plasma.
A one-dimensional version of the material was constructed as shown
in figure \ref{fig:diag}(a). It is a tube with an array of very thin
elastic membranes placed inside. The negative effective density has
been observed in the frequency range from 0 to 735 Hz, confirming
the theoretical prediction.

The metamaterial consists of the unit cells shown in figure
\ref{fig:diag}(a). A low density polyethylene membrane with
thickness of 0.01 mm  is placed inside a tube of 32.3 mm inner
diameter and 70 mm long. The membrane is stretched and the periphery
is attached leak-tight to the inner surface of the tube. The
resulting tension of the membrane is 65 N/m. When the fluid (air for
the present case) moves along the tube, the membrane is pushed to
form a paraboloid, and gives restoring force proportional to the
displacement of the fluid.

When the unit cells are connected, the one dimensional metamaterial
is formed as shown in figure \ref{fig:diag}(a). The acoustic wave in
the tube is described by the longitudinal displacement of the fluid,
$\vec{\xi}(x, t)$. If the length of the unit cell is much smaller
than the wavelength (the long-wavelength limit), the restoring
forces of the membranes generate static pressure gradient
proportional to $\vec{\xi}$,

\begin{equation} \label{eq:sound-1}
{\bigtriangledown}p = -\kappa  \vec{\xi}
\end{equation}
where $p$ and $\kappa$ are the pressure and the new elastic modulus,
respectively. In three dimensions, several types of membrane
structures are possible. The simple cubic type, for example,
consists of cubic cells with windows of thin elastic membranes on
all six walls. Generally speaking, the new elasticity can be
regarded as an intrinsic property that characterizes the
metamaterial according to equation (\ref{eq:sound-1}), independent
of the particular structure that generates it. Because the fluid is
elastically anchored in space by the membranes, we will refer to
this type of elasticity as 'spatially anchored elasticity' (SAE).

In the dynamic case, due to the force from the membranes, Newton's
equation becomes $- \bigtriangledown p = {\rho}' \partial
\vec{u}/\partial t + \kappa \vec{\xi}$, where $\vec{u} = \partial
\vec{\xi}/\partial t$ is the velocity of the fluid particle. The
volume-averaged density ${\rho}'$ of the fluid and the membranes,
giving the inertia term in the equation, is significantly different
from the dynamic effective mass density ${\rho}_{eff}$ defined
below. Using the harmonic expressions $\vec{u}(x,t) = \vec{U}
e^{-i\omega t}$, the equation can be written in a convenient form,
\begin{equation} \label{eq:sound-2}
-{\bigtriangledown}p = \left( {\rho}'-\frac{\kappa}{\omega
^2}\right) \frac{\partial \vec{u}}{\partial t} .
\end{equation}
Now the proportionality constant of the acceleration to the pressure
gradient force is defined as the effective density,
\begin{equation} \label{eq:sound-3}
\rho_{eff} = {\rho}' - \frac{\kappa}{\omega^2} = {\rho}' \left(
1-\frac{\omega^2_{SAE}}{\omega ^2}\right),
\end{equation}
which becomes negative below the critical frequency $\omega_{SAE} =
\sqrt{\kappa/{\rho}'}$  $\left( f = \omega/2 \pi \right)$.
Interestingly equation (\ref{eq:sound-3}) has essentially the same
form as the expression for the permittivity $\epsilon$ in a plasma,
with ${\omega}_{SAE}$ corresponding to the plasma frequency
${\omega}_{p}$. The acoustic wave equation obtained from equation
(\ref{eq:sound-2}) and the continuity equation $\bigtriangledown
\cdot \vec{u} = - (1/B)\partial p/\partial t$ give the
frequency-dependent phase velocity,
\begin{equation} \label{eq:sound-4}
v_{ph} = \sqrt{\frac{B}{\rho_{eff}}} = \sqrt{\frac{B}{{\rho}' \left(
1-{\omega^2_{SAE}}/{\omega ^2}\right)}}.
\end{equation}
For frequencies below ${\omega}_{SAE}$, the acoustic waves do not
propagate because the phase velocity assumes an imaginary value.
Above this frequency, the phase velocity assumes a large value and
decreases down to $\sqrt{{B}/{\rho}'}$  as the frequency becomes
high.

To experimentally verify this result, the pressure is measured in
each cell using the setup shown in figure \ref{fig:diag} (b). A
small sound source is placed at the beginning end of the
metamaterial, and the other end is terminated by an absorber to
prevent any reflection so that the acoustic wave propagating in the
metamaterial behaves as if the metamaterial extends to infinity. The
absorber is a long metamaterial of the same kind with dissipation
elements additionally placed in each cell. The dissipation element
is a sponge-like plate placed across the tube to generate a drag
force against the longitudinal motion of the fluid. Pressure
measurement confirmed that the sound decayed exponentially away as
it propagated in the absorber, and there was almost no reflection at
the metamaterial/absorber boundary (data not shown). This eliminates
concerns about the effect of the finite number of cells used in the
experiment, as well as the interference effect from the reflected
waves. For the phase velocity data, pressure was measured as a
function of time and position in the pass band. The detector was
inserted into the cell through the tube wall using feed-through
plugs and moved from cell to cell. The detector did not alter the
characteristics of the media.

The average density of the air loaded with the membrane in the tube
was ${\rho}' \sim 1.34$ kg/m$^3$, about 10\% higher than the density
of air which is ${\rho}_0 \sim 1.21$ kg/m$^3$ ~\cite{black}. The
modulus in equation (\ref{eq:sound-1}) was calculated from the
tension of the membranes to be $\kappa = 2.85 \times 10^7$ N/m$^4$.
From these values a critical frequency of $\omega_{SAE} =
\sqrt{\kappa/{\rho}'}$ was calculated to be about 735 Hz $(f =
{\omega}/2 {\pi})$.

\begin{figure}
\begin{center}
\includegraphics*[width=0.45\columnwidth]{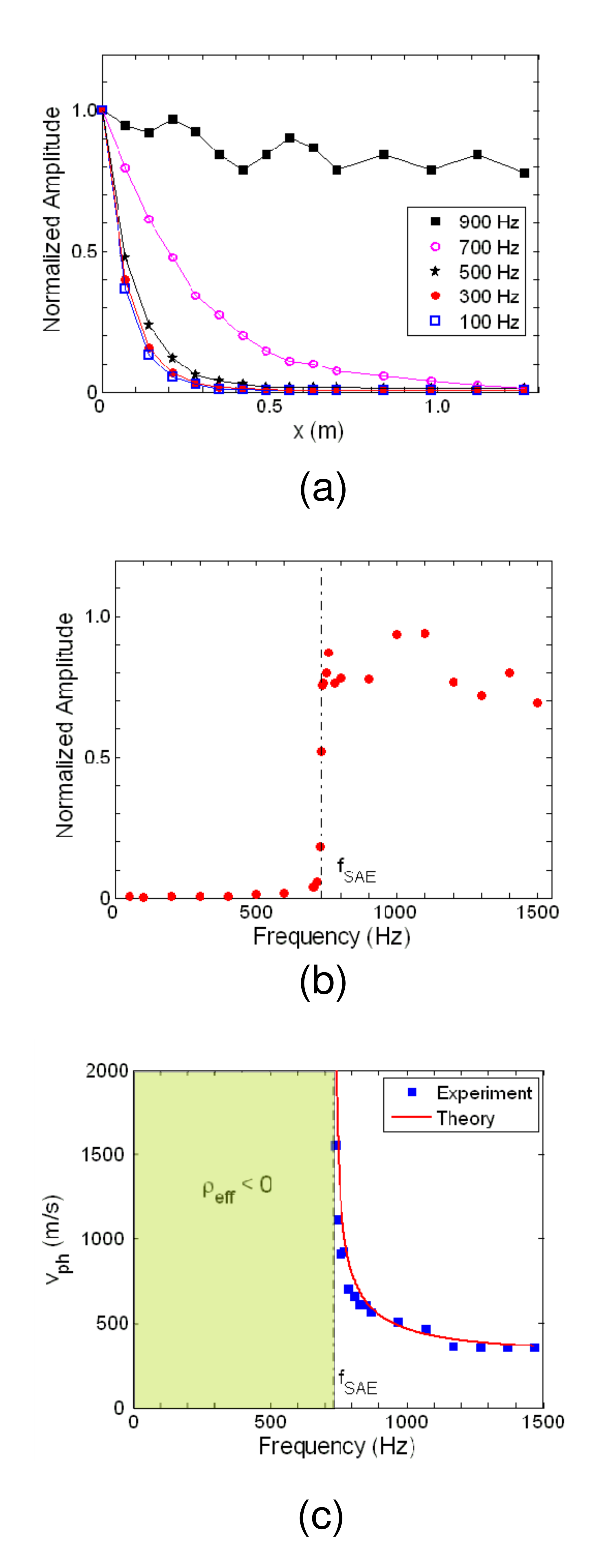}
\end{center}
\caption{(color online) (a) Sound intensities as functions of the
distance from the source, $x$. (b) Transmittance data in the
metamaterial. The cut-off frequency of the transmission data agrees
with the calculated value of $f_{SAE}$ (indicated with a broken
line). (c) Phase velocities in the metamaterial. } \label{fig:sae}
\end{figure}

Figure \ref{fig:sae} (a) shows propagation data of acoustic waves in
the metamaterial for several frequencies. For the frequencies below
$\omega_{SAE}$ (100 Hz, 300 Hz, 500 Hz, 700 Hz) the sound intensity
decayed exponentially with the distance, $x$, from the sound source.
Lines connecting data points are drawn for eye guides. For the
frequencies above $\omega_{SAE}$ (only 900 Hz is shown for clarity)
the sound waves propagated well without decay. The sound intensities
were normalized to the intensity level at $x = 0$. The transmissions
of the acoustic waves in the metamaterial were measured as the
ratios of pressure intensities at two positions ; at $ x = 0$ and $
x = 1.3$ m. The cut-off frequency of the transmission data agrees
with the calculated value of $\omega_{SAE}$ (indicated with a broken
line). Experimental and theoretical values of the phase velocities
are shown in figure \ref{fig:sae} (c). The theoretical phase
velocity predicted by equation (4) agrees excellently with the
experimental data.

Note that even though the SAE is a kind of elasticity, it has no
effect on the value of the effective modulus, but it strongly alters
the value of $\rho_{eff}$. The mechanism for the formation of
$\rho$-negativity from the SAE is different from that of
local-resonant type acoustic metamaterials~\cite{7,9}. In addition,
the SAE offers a substantially wider frequency range for negative
effective density.

In conclusion, we described the fabrication of a new class of
acoustic metamaterials in this paper. We introduced the novel
concept of 'spatially anchored elasticity' which uses a homogenized
structure of membranes to produce negative effective density. The
constructed structure exhibited negative effective density
characteristics in the spectral range from 0 to 735 Hz. Above this
frequency, the phase velocity is highly dispersive. Experimental
transmission and phase velocity measurement agree excellently with
the theoretical predictions. We expect the acoustic density
negativity presented in this paper to provide a useful basis for
applications in acoustic superlensing and
cloaking~\cite{aa1,aa2,aa3,aa4,aa5,aa6}.

\begin{acknowledgments}
The research was partially supported by The Korea Science and
Engineering Foundation (KOSEF R01-2006-000-10083-0).
\end{acknowledgments}


\end{document}